\documentclass[12pt]{article}
\usepackage{epsfig}

\def\beq{\begin{equation}}
\def\eeq#1{\label{#1}\end{equation}}
\def\eeqn{\end{equation}}

\def\beqa{\begin{eqnarray}}
\def\eeqa#1{\label{#1}\end{eqnarray}}
\def\eeqan{\end{eqnarray}}

\let\bar=\overbar

\def\Dslash{\not{\hbox{\kern-4pt $D$}}}
\def\dslash{\not{\hbox{\kern-2pt $\del$}}}

\def\msb{{\bar{\ssstyle M \kern -1pt S}}}

\def\KS {K_s}
\def\Bbar    {\kern 0.18em\overline{\kern -0.18em B}{}}
\def\Kbar    {\kern 0.18em\overline{\kern -0.18em K}{}}
\def\babar{\mbox{\sl B\hspace{-0.4em} {\small\sl A}\hspace{-0.37em} \sl B\hspace{-0.4em} {\small\sl A\hspace{-0.02em}R}}}

\def\Title#1{\begin{center} { {\bf #1} } \end{center}}

\begin{document}

\Title{\Large 
\mbox{\boldmath$B$}\kern-0.12em 
\large \mbox{\boldmath$A$}\kern-0.12em 
\Large \mbox{\boldmath$B$}\kern-0.12em 
\large \mbox{\boldmath$A\kern-0.12em R$}
\Large Results on \mbox{\boldmath$B \to X_s \gamma$}}

\bigskip\bigskip


\begin{raggedright}  

{\it Jack L. Ritchie\index{Reggiano, D.}\\
Department of Physics \\
University of Texas at Austin\\
Austin, TX  78712, USA \\
Representing the \babar\ Collaboration}
\bigskip\bigskip
\end{raggedright}

\noindent {Proceedings of CKM 2012, the 7th International Workshop on the CKM Unitarity 
Triangle, University of Cincinnati, USA, 28 September - 2 October 2012}

\section{Introduction}

\begin{figure}[htb]
\begin{center}
\epsfig{file=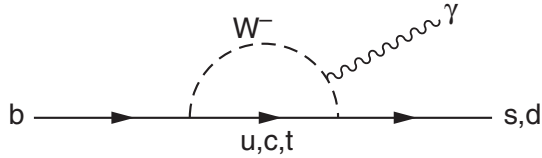,height=1.0in}
\caption{The electromagnetic penguin diagram
responsible for $B \to X_s \gamma$ decays in the SM.}
\label{fig:bsgdiag}
\end{center}
\end{figure}

The flavor-changing neutral current process $b \to s \gamma$, shown in
Figure~\ref{fig:bsgdiag}, is of
interest because it is one of the most reliably calculable of such processes
in the Standard Model (SM) and also because many new physics scenarios (e.g., SUSY)
may lead to deviations from the SM decay rate.  
Heavy-quark hadron duality implies
this decay rate is very close to the decay rate for $B \to X_s \gamma$, where 
$X_s$ represents any
hadronic system containing a strange particle.  The current next-to-next-to
leading order calculation \cite{Misiak} gives:
$${\cal B}(B \to X_s \gamma) = (3.15 \pm 0.23) \times 10^{-4},
$$
for $E_\gamma > 1.6 \, {\rm GeV}$ ($E_\gamma$ is the photon energy in the rest frame of the $B$ meson).  
Measurements of the photon spectrum can constrain Heavy Quark Effective Theory parameters and
help to reduce the uncertainties in the extraction of CKM elements $|V_{cb}|$ and $|V_{ub}|$
from semileptonic $B$ decays.

Experimentally, measuring ${\cal B}(B \to X_s \gamma)$ is challenging.  Multiple approaches are
undertaken, each with strengths and weaknesses.  \babar's final results are presented here for two of the 
alternative approaches.
In one, a fully inclusive measurement is 
performed by detecting only the high-energy
photon from the signal $B$ decay, and using a lepton ($e$ or $\mu$) from the semileptonic
decay of the other $B$ in the event to suppress backgrounds.  This method has the advantage of being inclusive, but the photon energy is smeared by the energy resolution of the electromagnetic calorimeter and also by the
motion of the signal $B$ in the $\Upsilon(4S)$ center of mass frame. A subtraction of backgrounds from other $B$ decays ultimately leads to a systematic error that is larger than the statistical uncertainty.  Also, this method does not distinguish $B \to X_s \gamma$ events from
$B \to X_d \gamma$, so effectively they are combined.
The second approach is a semi-inclusive measurement, 
in which a large 
number of exclusive modes are fully reconstructed and combined.  This method has the virtue
that the signal $B$ is reconstructed, providing a precise determination of the photon energy 
in the $B$ rest frame via the relation
$E_\gamma = (m_B^2 - m_X^2)/(2m_B)$, but suffers from the fact that the measurement 
is not inclusive.  Many modes are not included, and
the uncertainty in estimating the missing modes introduces a systematic error that dominates the measurement.
In both cases, all analysis procedures and event selection criteria
were determined before they were applied to real data in the signal regions.
 
\section{Fully Inclusive Results}

\begin{figure}[hbt]
\begin{center}
\epsfig{file=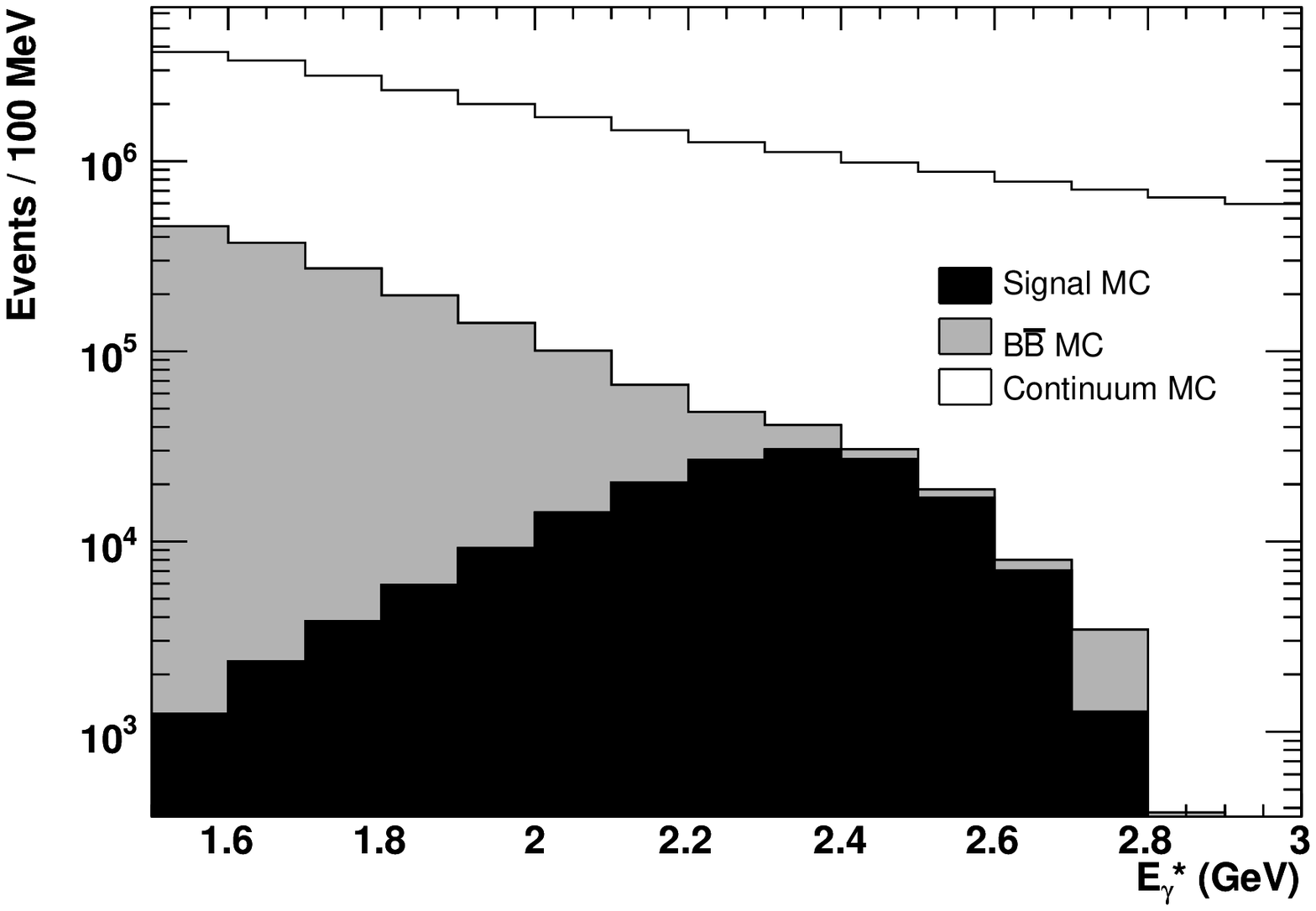,height=2.0in}
\epsfig{file=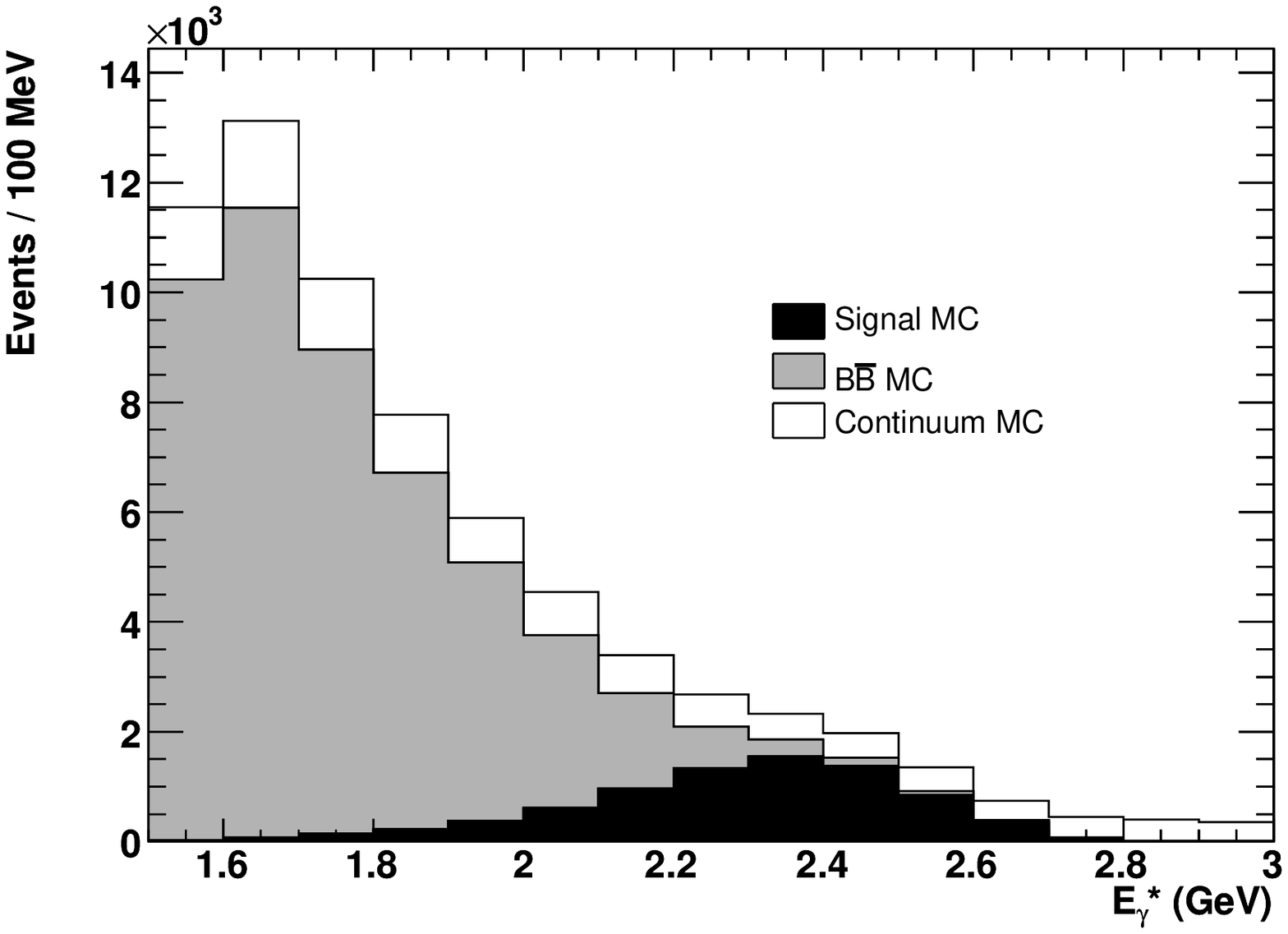,height=2.0in}
\caption{Monte Carlo signal and background yields versus $E_\gamma^*$: (a) after selecting
a high-energy photon (log scale) and (b) after all selection criteria (linear scale).}
\label{fig:MCspectra}
\end{center}
\end{figure}

The fully inclusive analysis \cite{inclusive-PRL, inclusive-PRD}
used $347 \,{\rm fb^{-1}}$ of data collected at the $\Upsilon(4S)$ at the SLAC PEP-II B-factory.  
The challenge is illustrated in Figure~\ref{fig:MCspectra}(a), which shows the expected SM signal, the $B$ decay 
background, and the continuum background after selection of a high-energy photon satisfying
$1.53 < E_\gamma^* < 3.5 \, {\rm GeV}$, where $E_\gamma^*$ is the photon energy in the $\Upsilon(4S)$ rest frame.
Figure~\ref{fig:MCspectra}(b) shows the situation
after all event selection criteria have been applied (achieving background rejection of about $10^{-5}$ with
a signal efficiency of 2.6\%).  
The large $B \Bbar$ background can only be suppressed by imposing a cut on 
the photon energy, chosen to be $E_\gamma^* > 1.8 \, {\rm GeV}$.  

\begin{figure}[htb]
\begin{center}
\epsfig{file=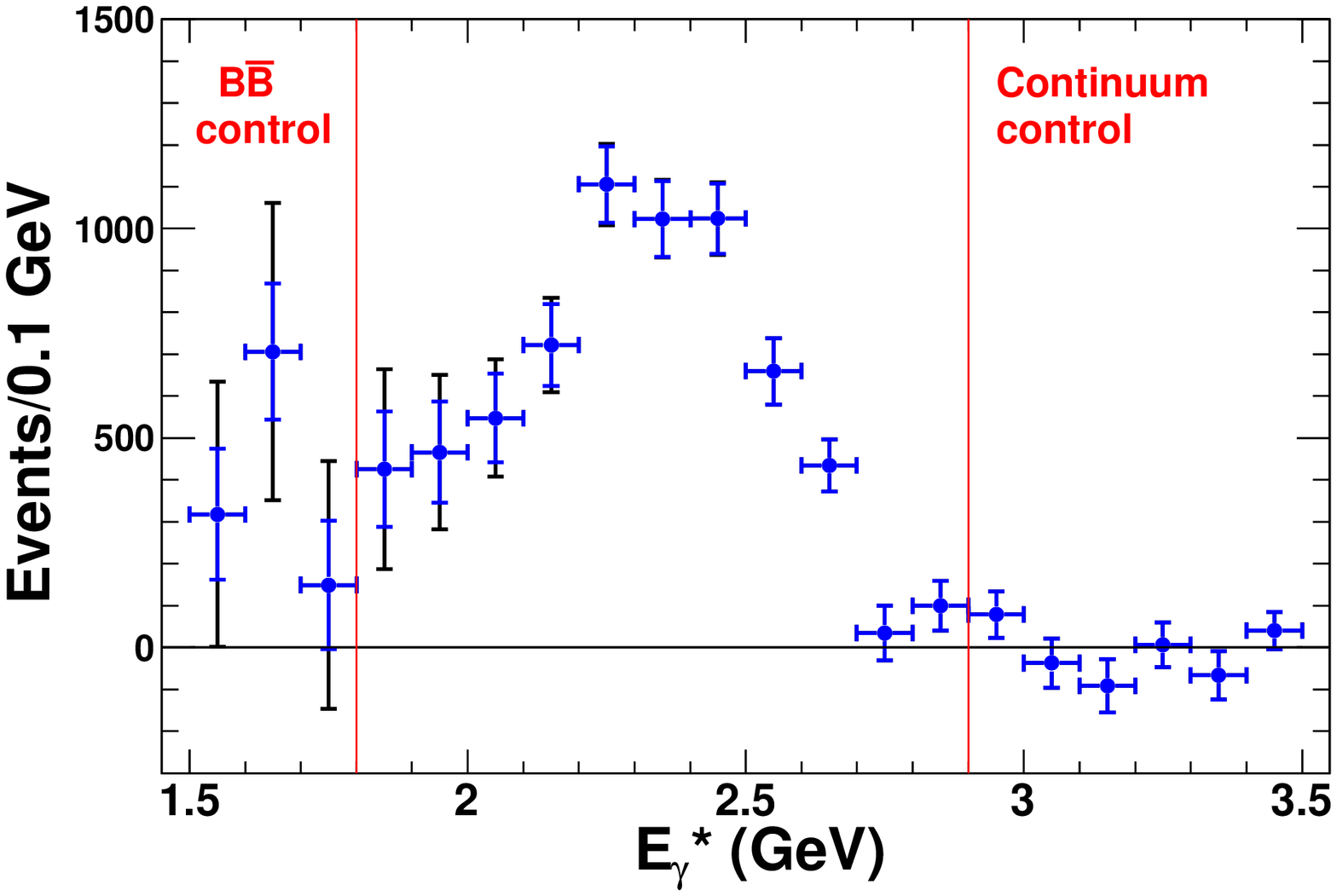,height=2.0in}
\epsfig{file=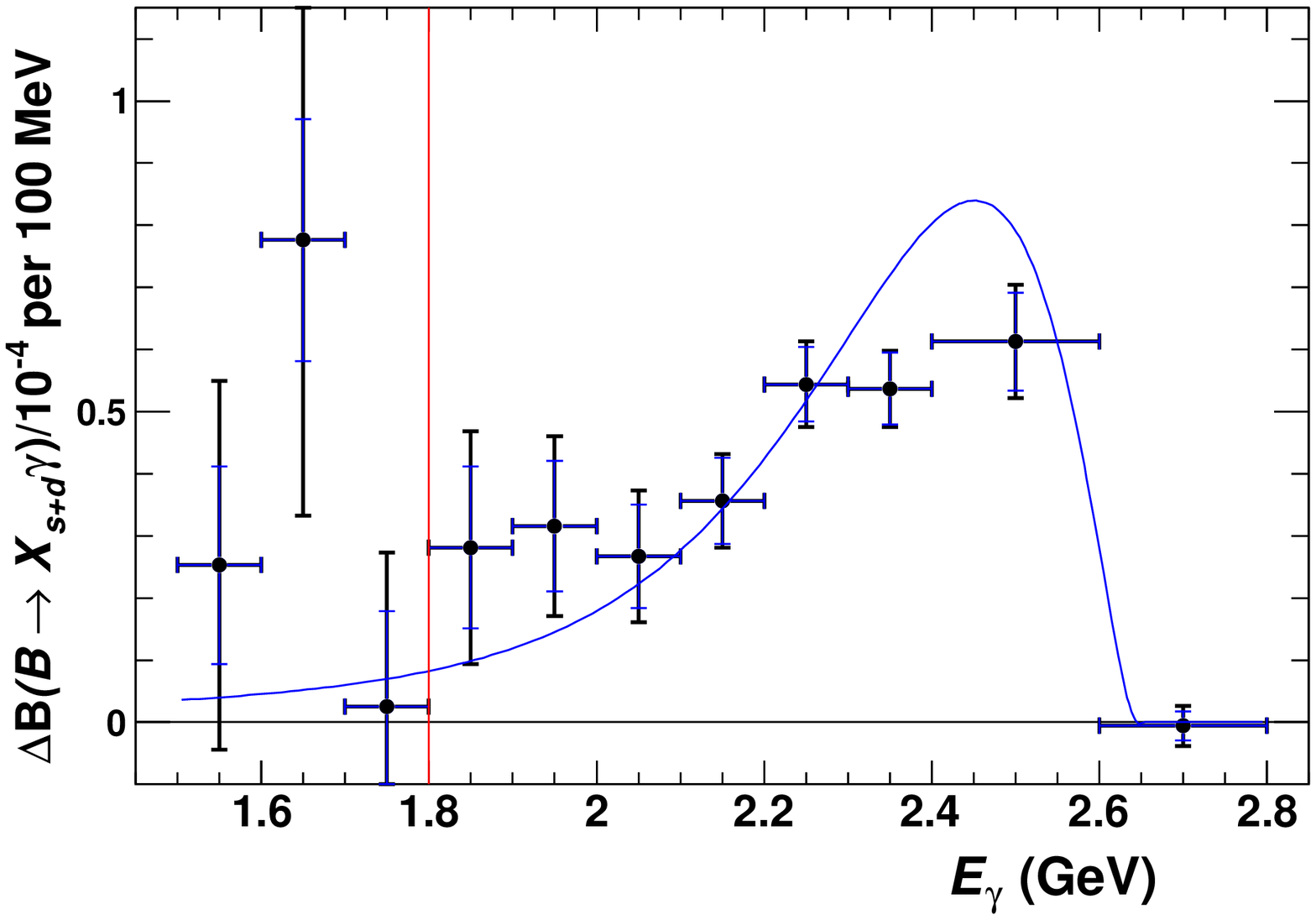,height=2.0in}
\begin{picture}(0,0)
\put(-150.0,140.0){(a)}
\put(90.0,140.0){(b)}
\end{picture}
\caption{(a) Photon spectrum after background subtraction.  The vertical lines indicate the boundaries
of the blind signal region. Inner error bars indicate statistical errors only.
(b) Photon spectrum after unfolding.  The curve shows the spectrum in the kinetic
scheme model using HFAG world average parameters normalized
to data in the range $1.8 < E_\gamma^* < 2.8 \, {\rm GeV}$.}
\label{fig:inclusive-spectra}
\end{center}
\end{figure}

The remaining backgrounds are subtracted using Monte Carlo that is corrected using data control samples.  The result for data, after $B \Bbar$ background subtraction,
is shown in Figure~\ref{fig:inclusive-spectra}(a).

The branching fraction is extracted from the event yield in the signal region defined by 
$1.8<E_\gamma^*< 2.8 \, {\rm GeV}$ by applying a signal efficiency correction, corrections for
smearing due to calorimeter energy resolution and motion of the signal $B$ in the $\Upsilon(4S)$ frame, and
a correction to account for $B \to X_d \gamma$ contamination (assuming the rates are related by
CKM factors and using $1/(1+|V_{td}/V_{ts}|^2)=0.958 \pm 0.003)$.  The result is
$$
{\cal B}(B \to X_s \gamma) = (3.21 \pm 0.15 \pm 0.29 \pm 0.08) \times 10^{-4},
$$
for $E_\gamma > 1.8 \, {\rm GeV}$, where the first error is statistical, the second is systematic, and the third represents the uncertainty in the signal efficiency from uncertainties in the model for the photon spectrum. 

An unfolding procedure is applied to extract the photon spectrum in the $B$ meson rest frame.  This procedure
corrects for calorimeter resolution smearing, smearing due to the motion of the $B$ in the
$\Upsilon(4S)$ rest frame, and corrects for detector and selection efficiencies.  The unfolded spectrum
is shown in Figure~\ref{fig:inclusive-spectra}(b).  In addition, the first, second, and third moments of
this spectrum, useful for determining Heavy Quark Effective Theory (HQET) model parameters, are reported in Reference~\cite{inclusive-PRD}.

A test for direct CP violation can be made by comparing the decay rates for $B$ versus $\Bbar$, using the charge
of the tag lepton to separate the two categories.  This measurement, which includes $B \to X_d \gamma$ events,
is a strong test for new physics since the SM expectation\cite{Benzky} is negligibly different from zero. 
The result, in the optimized energy range $ 2.1 < E_\gamma^* < 2.8 \, {\rm GeV}$, is
$$A_{CP} = 
{{\Gamma(B \to X_{s+d}\gamma) - \Gamma(\Bbar \to X_{s+d}\gamma)} \over 
{\Gamma(B \to X_{s+d}\gamma) + \Gamma(\Bbar \to X_{s+d}\gamma)}} 
= 0.057 \pm 0.060 \pm 0.018.
$$

\section{Semi-inclusive Results}

\begin{figure}[htb]
\begin{center}
\epsfig{file=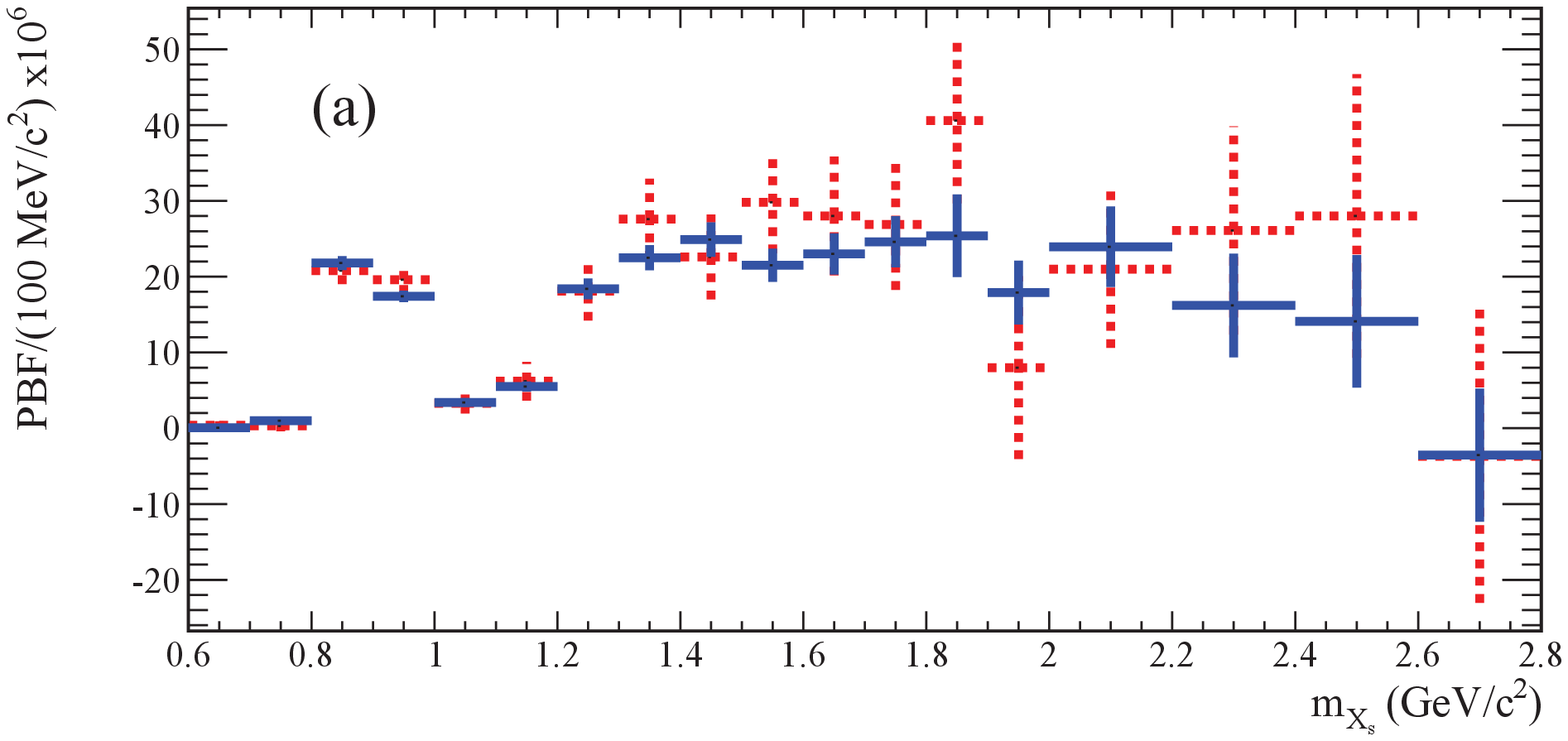,height=1.8in}
\epsfig{file=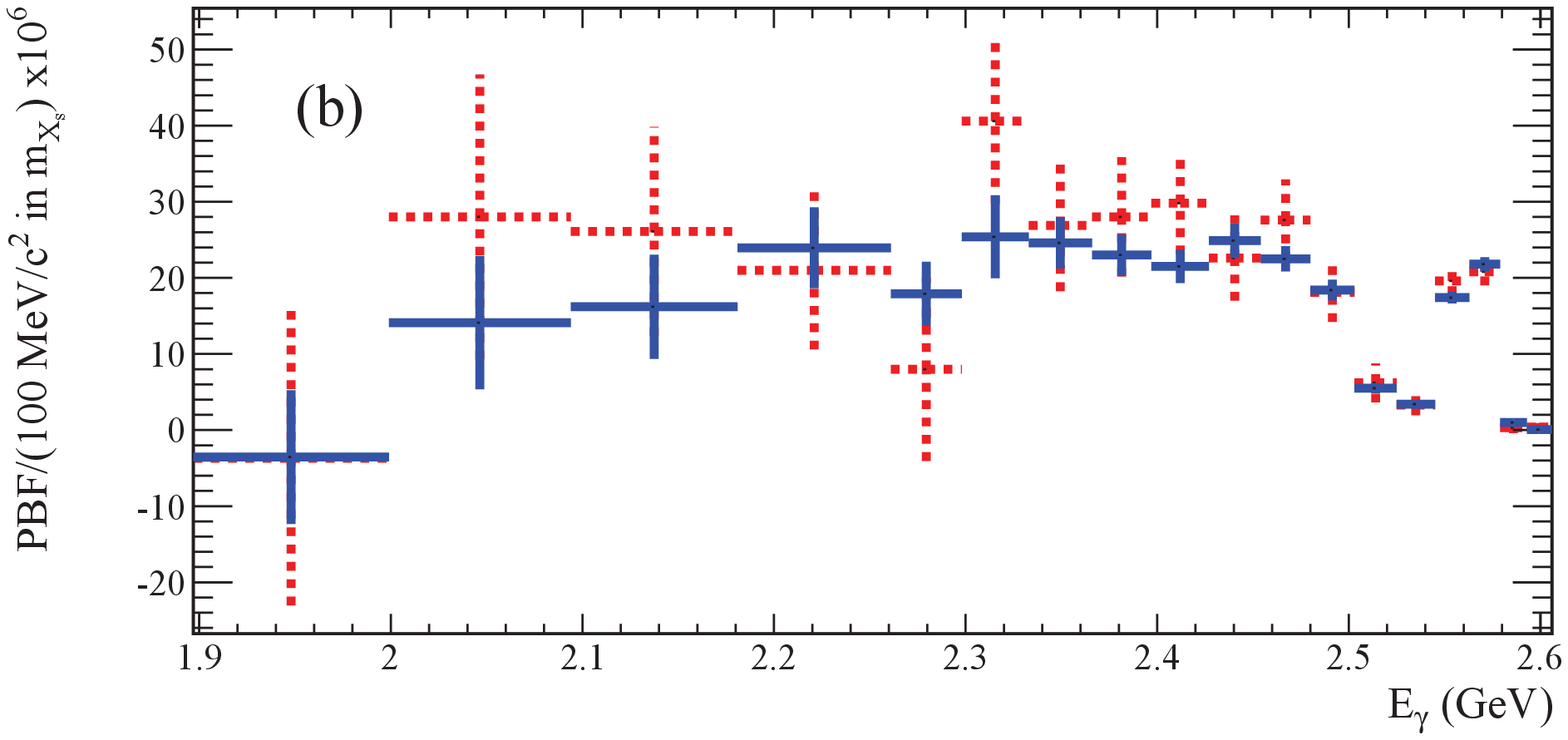,height=1.8in}
\caption{(a) Reconstructed hadronic mass spectrum and (b) the associated photon spectrum.
The solid (blue) data points correspond to this measurement.  The dashed (red) points are from
an earlier \babar\ measurement with $82 \, {\rm fb^{-1}}$.}
\label{fig:semispectra}
\end{center}
\end{figure}

The semi-inclusive analysis \cite{semi} used $429 \,{\rm fb^{-1}}$ of data 
and reconstructs the 38 exclusive modes listed
in Table~\ref{tab:38modes}.  
\begin{table}[hb]
\begin{center}
\begin{tabular}{l l l l}  
\hline
$ \KS\pi^{+}$ \qquad & $\KS\pi^{+}\pi^{-}\pi^{+}$ \qquad 
                     & $ K^{+}\pi^{+}\pi^{-}\pi^{-}\pi^{0}$ \qquad\qquad & $ \KS\eta\pi^{+}\pi^{-}$ \\
$ K^{+}\pi^{0}$ & $ K^{+}\pi^{+}\pi^{-}\pi^{0}$ 
                     & $ \KS\pi^{+}\pi^{-}\pi^{0}\pi^{0}$ & $ K^{+}\eta\pi^{-}\pi^{0}$ \\
$ K^{+}\pi^{-}$ & $ \KS\pi^{+}\pi^{0}\pi^{0}$
                     & $ K^{+}\eta$ & $ K^{+}K^{-}K^{+}$ \\
$ \KS\pi^{0}$ & $ K^{+}\pi^{+}\pi^{-}\pi^{-}$	
                     & $ \KS\eta$ & $ K^{+}K^{-}\KS$ \\
$ K^{+}\pi^{+}\pi^{-}$ \qquad \qquad & $ \KS\pi^{0}\pi^{+}\pi^{-}$ 
                     & $ \KS\eta\pi^{+}$ & $ K^{+}K^{-}\KS\pi^{+}$ \\
$ \KS\pi^{+}\pi^{0}$ & $ K^{+}\pi^{-}\pi^{0}\pi^{0}$ 
                     & $ K^{+}\eta\pi^{0}$ & $ K^{+}K^{-}K^{+}\pi^{0}$  \\
$ K^{+}\pi^{0}\pi^{0}$ & $ K^{+}\pi^{+}\pi^{-}\pi^{+}\pi^{-}$ \qquad\qquad
                     & $ K^{+}\eta\pi^{-}$ & $ K^{+}K^{-}K^{+}\pi^{-}$ \\
$ \KS\pi^{+}\pi^{-}$ & $ \KS\pi^{+}\pi^{-}\pi^{+}\pi^{0}$
                     & $ \KS\eta\pi^{0}$ & $ K^{+}K^{-}\KS\pi^{0}$ \\
$ K^{+}\pi^{-}\pi^{0}$ & $ K^{+}\pi^{+}\pi^{-}\pi^{0}\pi^{0}$
                     & $ K^{+}\eta\pi^{+}\pi^{-}$ & \\
$ \KS\pi^{0}\pi^{0}$ & $ \KS\pi^{+}\pi^{-}\pi^{+}\pi^{-}$
                     & $ \KS\eta\pi^{+}\pi^{0}$ & \\
\hline
\end{tabular}
\caption{38 final states reconstructed, with $\KS \to \pi^+ \pi^-$, 
$\pi^0 \to \gamma \gamma$, and $\eta \to \gamma \gamma$.}
\label{tab:38modes}
\end{center}
\end{table}
Signal selection and background rejection is accomplished using
random forest classifiers trained on Monte Carlo signal and background samples.  After 
reconstruction, events are binned as a function of the mass of the hadronic system, $m_X$,
and maximum likelihood fits are performed in each of 18 mass bins to extract signal yields.
Figure~\ref{fig:semispectra}(a) shows the resulting $m_X$ spectrum.  The corresponding
photon spectrum, shown in Figure~\ref{fig:semispectra}(b), is obtained using 
$E_\gamma = (m_B^2 - m_X^2)/(2m_B)$.

The branching fraction is obtained summing the partial (bin-by-bin) branching fractions
shown in Figure~\ref{fig:semispectra}, giving the result:
$${\cal B}(B \to X_s \gamma) = (3.29 \pm 0.19 \pm 0.48) \times 10^{-4},
$$
for $E_\gamma > 1.9 \, {\rm GeV}$.  The first error is statistical and the second error
is systematic.  The systematic error, which dominates, is mainly due to the missing modes.

Moments of the spectrum are also measured and the spectrum is fit to determine
the parameters of HQET models.  In particular, fits are performed
to determine the parameters $m_b$ and $\mu_b^2$ in both the ``kinetic" scheme and the ``shape function"
scheme.  The results are given in Table~\ref{tab:HQET}.

\begin{table}[htb]
\begin{center}
\begin{tabular}{l|ll}  
 &  Kinetic scheme &  Shape function scheme \\ \hline
 $m_b$  &   $4.568^{+0.038}_{-0.036} \, {\rm GeV/c^2}$    &     $4.579^{+0.032}_{-0.029} \, {\rm GeV/c^2}$     \\
 $\mu_b^2$ &  $0.450 \pm 0.054 \, {\rm GeV^2}$     &     $0.257^{+0.034}_{-0.039} \, {\rm GeV^2}$    \\ \hline
\end{tabular}
\caption{Results for HQET parameters based on fits to the $m_X$ spectrum.}
\label{tab:HQET}
\end{center}
\end{table}

\section{Conclusions}

\begin{figure}[htb]
\begin{center}
\epsfig{file=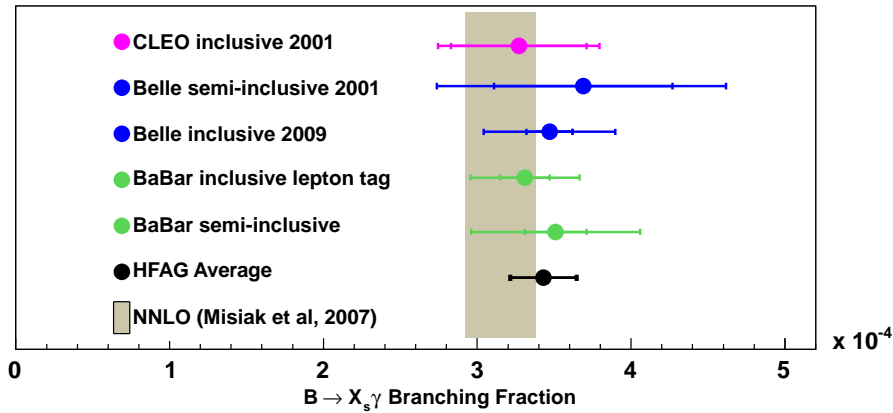,height=2.25in}
\caption{Summary of $B \to X_s \gamma$ branching fraction measurements: CLEO inclusive\cite{CLEO}, Belle semi-inclusive\cite{Belle-semi}, Belle inclusive\cite{Belle-inclusive},
and the \babar\ results presented here for inclusive\cite{inclusive-PRL, inclusive-PRD} and 
semi-inclusive\cite{semi}. The shaded region shows the SM theory expectation\cite{Misiak}. 
The current HFAG average\cite{HFAG} is also shown.}
\label{fig:BFsummary}
\end{center}
\end{figure}

\babar\ has reported final results from two analyses of $B \to X_s \gamma$,
providing branching fractions, spectra, and spectral moments, as well as testing for
direct CP violation.  The branching
fraction measurements are in good agreement with the SM theory expectation, as shown in
Figure~\ref{fig:BFsummary}.  The figure shows comparable results of all experiments, extrapolated to
a common energy cutoff of $1.6 \, {\rm GeV}$ using the Heavy Flavor Averaging Group (HFAG)
extrapolation factors \cite{HFAG}.

\noindent This work was supported in part by Department of Energy contract DE-AC02-76SF00515.


%
%
%
%
 
\end{document}